\documentstyle[prl,aps,psfig,floats,twocolumn]{revtex} 
\begin{document} 
\draft 
\twocolumn[\hsize\textwidth\columnwidth\hsize\csname 
@twocolumnfalse\endcsname 
 
\title{A perturbative approach to the Bak-Sneppen Model} 
\author{M. Felici$^{1,2}$,  G. Caldarelli$^{1}$, A. Gabrielli$^{1,2}$,  
and L. Pietronero$^{1}$} 
\address{$^1$ INFM - Unit\`a di Roma1 and Dip. Fisica, Universit\`a 
``La Sapienza'' P.le A. Moro 2, 
00185 - Roma, Italy}
\address{$^2$ Laboratoire de la Physique de la Mati\`ere Condens\'ee, 
Ecole Polytechnique, 91128 Palaiseau, France} 
 
\date{\today} 
\maketitle 
\begin{abstract} 
We study the Bak-Sneppen model in the probabilistic framework of the 
Run Time Statistics (RTS). 
This model has attracted a large interest for its simplicity  
being a prototype for the whole class of models showing Self-Organized  
Criticality. 
The dynamics is characterized by a self-organization of almost 
all the species fitnesses  
above a non-trivial threshold value, and by a lack of spatial and temporal 
characteristic scales. This results in {\em avalanches} of activity
power law distributed.  
In this letter we use the RTS  approach to compute the value of $x_c$, 
the value of 
the avalanche exponent $\tau$ and the asymptotic distribution of minimal
fitnesses.
\end{abstract} 
\pacs{02.50.-r, 64.60.Ak, 04.20.Jb} 
] 
\narrowtext 
 
The concept of Self-Organized Criticality (SOC) has been introduced  
in order to explain the ubiquitous presence of scale free phenomena in 
Nature. Ranging from fractures\cite{frac1} 
to  river basins\cite{rinaldo}, the lack of spatial and/or 
temporal characteristic scales is the 
main feature of these processes. 
To detect if an underlying general mechanism exists, 
two paradigmatic models have been introduced. 
One mimics the evolution of sandpiles \cite{btw}, and the  
other one, has been originally introduced to  
mimic punctuated equilibrium in the evolution of the ecology of species. 
We focus here on the latter one, known as Bak-Sneppen (BS)  
model\cite{bs}. 
We develop a perturbative approach to BS based on the Run Time 
Statistics \cite{luciano,matteo}.
This allows us to evaluate the main quantities for the 
model in the stationary state: the fitnesses histogram, the distribution of 
minimal fitnesses, and the avalanche exponent. 

The BS model is defined as follows: 
an $1$-d ring of $N$ different sites models the $N$ species present in the ecosystem. 
Links between first nearest neighbor (fnn) sites model interaction 
between species (i.e. predation).  
A random number $x_i \in [0,1]$, extracted from the uniform 
probability density function (pdf) $f_0(x)=1$, is assigned 
to each site/species $i$ in the system.  
The $x_i$'s quantify the fitness of species $i$ to survive the competition 
with the other species in the environment.  
At each time step, the species with the minimal fitness is selected 
and removed. 
The species connected to it through predation (i.e. the fnn) 
are also removed, regardless their fitness. 
Three new species with randomly extracted fitnesses take their place.
We refer to this updating rule of the model as the {\em refreshing} rule. 
After a transient period this system self-organizes in a stationary state 
characterized by two main features: (1) the normalized fitnesses histogram
$\phi(x)$ is given, apart from corrections of order $1/N$,
by $\phi(x)=\theta(x-x_c)/(1-x_c)$,
where $\theta(x)$ is the usual step function, and $x_c=0.66702(3)$ \cite{bsdata};
(2) the stationary dynamics evolves as a sequence of {\em critical avalanches} 
\cite{bs,matteobs,rev}, the duration $s$ of which is power law
distributed: 
$P(s)\sim s^{-\tau}$ with $\tau\simeq 1.07$.

The best analytical results for BS come from mean field approaches 
giving a threshold $x_c=1/3$ \cite{derr}, very far from the real 
numerical value. 
This striking disagreement lead us to investigate if the actual value can  
be computed by using a theoretical tool, called Run Time Statistics 
(RTS) \cite{luciano,matteo}. RTS is designed to study growth processes in  
a medium with quenched disorder. 
In particular, the main results \cite{euro1,frac} obtained through RTS 
deal with the class of models derived by the Invasion Percolation (IP) 
in $d=2$\cite{wilkinson}, to which the BS model can be related. 

RTS is based on the intuitive idea that the
larger the number of time-steps one species survives,
then {\em probably} the larger its fitness is. 
In other words, if at the first time-step with no information, one assigns a 
uniform pdf for the species fitnesses, at successive time-steps
information about the history increases the
effective pdf at high fitness values for the surviving species.
More formally,
this information is stored in {\em effective} (conditional) pdf's 
$\{f_{i,t}(x)\}$ of the quenched disorder variables $\{x_i\}$.
These {\em effective} pdf's can then be used to develop
a {\em step by step} algorithm to describe probabilistically 
the whole tree of the possible dynamics of 
the system, starting from the last available situation \cite{matteo}.  
Let us suppose to know at time $t$ the set of effective pdf's 
$\{f_{i,t}(x)\}$.
By using joint probability, it is possible to write for each species
the effective probability to be selected at that time-step, 
and one can update the effective
pdf's, obtaining the new set of functions $\{f_{i,t+1}(x)\}$.
For example, consider a limit case with a system composed only by 
two sites $a, b$. The minimum
is selected and refreshed without affecting the other.
At the beginning ($t=0$) both pdf's are uniform and equal to $1$, and 
the probability $\mu_{a,0},\mu_{b,0}$ to be selected for 
sites $a, b$ are both equal to $1/2$.
We assume without loss of generality that $x_a<x_b$.
By using the postulate of the theory of conditioned
probability 
\begin{eqnarray}
f_{a,1}(x_a)&=&1 \nonumber \\
f_{b,1}(x_b)&=&\frac{f_{b,0}(x_b)}{\mu_{a,0}}\int_0^{x_b}dx_a f_{a,0}(x_a)=2x_b
\end{eqnarray}
and consequently 
\begin{eqnarray}
\mu_{a,1}&=&\int_0^1dx_b f_{b,1}(x_b)\int_{0}^{x_a} dx_a f_{a,1}(x_a)=2/3
\nonumber \\
\mu_{b,1}&=&\int_0^1dx_a f_{a,1}(x_a)\int_{0}^{x_a} dx_b f_{b,1}(x_b)=1/3.
\end{eqnarray}
That is the just refreshed site has a larger probability to grow.
This argument can be easily generalized for any time $t$. 
For instance, if the site selected is always the same, $f_{b,t}=(t+1)(x_b)^t$ and
$\mu_{b,t}=\frac 1 {t+2} \simeq \frac 1 t$.  

\begin{figure} 
\centerline{\psfig{file=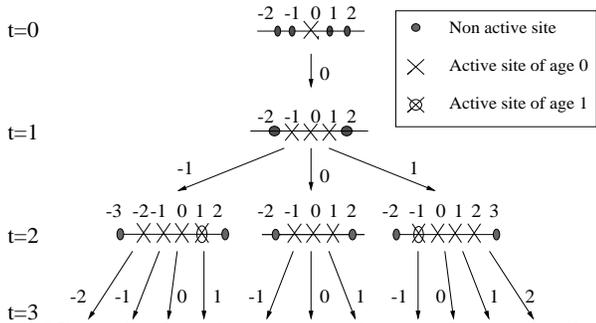,height=4.2cm,width=8.5cm,angle=0}} 
\caption{the first three temporal steps of an avalanche tree. 
The initiator $i=0$ is selected at $t=0$ and the possible growths are shown.} 
\label{fig1} 
\end{figure}

In order to implement this program for the BS, one needs only an initial site 
configuration with known $\{f_{i,t}(x)\}$.
There are two possibilities: 
(i) to follow the dynamics from the first time step $t=0$, having obviously
$f_{i,0}(x)\equiv f_0(x)=1$ for each $i$; 
(ii) to start at an arbitrary time $t$ for which the set 
$\{f_{i,t}(x)\}$ is known.
In the following, we consider mainly the case (ii):
the system is considered 
already at the critical stationary state, just after the first 
step of a critical avalanche.
We call the site selected at this first step the {\em initiator} of the 
avalanche.
We recall that the initiator fitness \cite{rev}
lies in an interval of order $1/N$ around $x_c$, and all the other
sites have a fitness $x>x_c$.
Avalanches are defined as geometrically and causally connected  
sequences of growths.  
Since it has been proved that each avalanche  
is independent on the others, the probabilistic study of  
a generic critical avalanche provides a complete statistical description of
the stationary state.
Without loss of generality, we consider the initiator 
placed at $i=0$ and its selection at $t=0$.
At $t=1$, the just grown initiator and its two fnn sites 
have the fitness uniformly
distributed because of the refreshing rule 
(i.e. $f_{0,1}(x)=f_{-1,1}(x)=f_{1,1}(x) =f_0(x)=1$).
Let us call $A_t$ ({\em active} sites) the set of sites refreshed  
by the avalanche up to its $t^{th}$ step. 
Consequently, $A_1$ is composed by the three sites $-1,0,+1$. 
The set $A_t$ defines a connected segment on the system 
because of the geometrical connectivity of an avalanche.   
One can represent 
the evolution of the system as a branching process (see Fig.~\ref{fig1}).  
At each time-step, the growth of each active site corresponds to a 
branching event.
The growth of ``non-active'' sites  
are not included in the branching tree, because
this implies the end of the avalanche and the beginning of a new one. 
Any finite connected path on this branching tree represents a 
single realization of the avalanche.  The tree contains all the possible  
realizations of the avalanche with a given initiator.

The RTS approach to an avalanche in the critical stationary state can be
formulated as follows.
The avalanche proceeds, after the selection of the initiator, if at least 
one of the three sites belonging to $A_1$ has fitness below $x_c$.  
In general, at the $t^{th}$ time-step of the avalanche, if one knows 
the effective pdf  $f_{i,t}(x)$ for each site $i \in A_t$, 
one can write the effective growth probability of site $i$  
conditioned to the fixed past history: 
\begin{equation} 
\mu_{i,t}=\int_0^{x_c}dx_i f_{i,t}(x_i)\prod_{j\ne i}\left[\int_{x_i}^1 
dy f_{j,t}(y) \right] 
\label{prob} 
\end{equation} 
Each of these possible growth events corresponds to a branch of  
the tree leading from the starting configuration 
to a new one characterized  
by a new set $A_{t+1}$ of active sites with their new effective 
pdf's.  
The set of sites $A_{t+1}$ is related with the ``mother'' set $A_t$.
If the just  
grown site $i$ is not an extreme of the segment defined by $A_t$, then
$A_{t+1}\equiv A_t$.
If, instead, $i$ is the left (right) extreme of $A_t$, then the left (right)
fnn of $i$ must be added to $A_t$ in order to obtain $A_{t+1}$.
After the selection of the site $i$, 
the new effective pdf's of the sites belonging to $A_{t+1}$ 
are obtained from those of $A_t$ by applying both the refreshing 
rule of the dynamics and the law of   
conditional probability $ P(B|A)=\frac{P(A\cap B)}{P(A)}$  
to store the information about the last growth event: 
\begin{eqnarray} 
f_{k,t\!+\!1}(x)\!&=&\left\{  
\begin{array}{rr} 
\!\!\frac{f_{k,t}(x)}{\mu_{i,t}}\!\int_0^x\!\!dx_if_{i,t}(\!x_i\!)\prod_{j}[ 
\int_{x_i}^{1}dy f_{j,t}(y)] \;\; x\leq x_{c}\\  
f_{k,t}(x_{c}) \qquad\qquad\qquad x>x_c\; 
\end{array} 
\right. \nonumber
\\ 
f_{i,t\!+\!1}(x)\!&=&\!\!f_{i-1,t\!+\!1}(x)=f_{i+1,t\!+\!1}(x)=f_0(x)=1 
\label{refresh2} 
\end{eqnarray} 
where $k$ is any species in the set $A_t$ 
different from $i$ and its neighbors, and $j$ is an active site different  
from $k,i$.  
The pdf of an active site is always constant above  
$x_c$ because in the stationary state no site with $x>x_c$ 
grows, then no information is available for 
$x>x_c$.  
Through this algorithm one can store step by step 
information on the fitnesses from the actual fixed history of the avalanche. 
By iterating Eqs.(\ref{prob},\ref{refresh2}), 
one can write for a time-length $\Delta t$ the probability of any connected path  
${\cal C}(\Delta t)$ 
belonging to the tree as  
\begin{equation} 
W({\cal C}(\Delta t))=\prod_{t=0}^{\Delta t} \mu_{i_t,t}\, ,
\label{weight} 
\end{equation} 
where $i_t$ is a site growing at time $t^{th}$ of the path ${\cal C}(\Delta t)$.

Let us now write an equation for the average fitness histogram $\phi_t(x)$.
From the refreshing rule, one has
\begin{equation} 
\begin{array}{ll}
N\phi_{t+1}(x)=N\phi_t(x)- m_{i,t+1}(x) - \hat{f}_{i-1,t+1}(x)\\ 
- \hat{f}_{i+1,t+1}(x)+3\, ,
\end{array}
\label{bala} 
\end{equation} 
where $m_{i,t+1}(x)$, $\hat{f}_{i-1,t+1}(x)$ and $\hat{f}_{i+1,t}(x)$ are 
the effective pdf's
that sites $i,i\!-\!1,i\!+\!1$ would have respectively after the selection
of site $i$, if their fitnesses were not refreshed.
Eq.(\ref{bala})  states that three species  are removed with their  
pdf's, and three new species with uniform pdf enter.
Considering the average $\left<...\right>$ of Eq.(\ref{bala})
over all the possible paths, and imposing
the stationarity condition $\phi_{t+1}(x)=\phi_t(x)=\phi(x)$, one has: 
\begin{equation} 
\left<m_{i}(x)\right>+\left<\hat{f}_{i-1}(x)+\hat{f}_{i+1}(x)\right>=3 \,,
\label{bala2} 
\end{equation} 
We use Eq.(\ref{bala2}) to estimate $x_c$
and the behavior of $\phi(x)$ under the simple hypothesis that
all the sites not touched by the avalanche have a fitness $x>x_c$ (which comes from
the stationarity condition).
The functions $\hat{f}_{i-1,t+1}(x)$ and $\hat{f}_{i+1,t+1}(x)$
are given by the second line of Eq.(\ref{refresh2}) with $k=i-1$ and  
$k=i+1$ respectively. 
In order to perform the average $\left<...\right>$, one has to consider all
the ways in which the selection of site $i$ represents  
the $t^{th}$ step of an avalanche (i.e. one has to consider all the
finite paths of the tree of Fig.~\ref{fig1}).
Then one can write:
\begin{eqnarray}
\label{secpezz}
&&\left<  
\hat{f}_{i-1}(x)+\hat{f}_{i+1}(x)\right>\\ 
&&=\frac{\sum_ 
t\sum_{{\cal C}(t)}  
W({\cal C}(t))\left[\hat{f}_{i-1,t+1}(x)+\hat{f}_{i+1,t+1}(x)\right]_{{\cal C}(t)}}  
{\sum_t\sum_{{\cal C}(t)} W({\cal C}(t))}\,, 
\nonumber
\end{eqnarray} 
where $W({\cal C}(t))$ is the RTS statistical weight of the generic 
path ${\cal C}(t)$ of length $t$, in the tree of Fig.~\ref{fig1}, 
given by Eq.(\ref{weight}).
Moreover $[..]_{{\cal C}(t)}$ indicates that the functions inside brackets
are evaluated by applying step by step the RTS algorithm to the path ${\cal C}(t)$. 
Note, anyway, that if $i$ is the left (right) extreme of $A_t$, 
then the knowledge of the path does not give any information about
the sites $i\!-\!1$ ($i\!+\!1$) apart that, before refreshment,
$x_{i-1}>x_c$ ($x_{i+1}>x_c$). In this case we can approximate 
$[\hat{f}_{i-1,t+1}(x)]_{{\cal C}(t)}=\phi(x)$ 
($[\hat{f}_{i+1,t+1}(x)]_{{\cal C}(t)}=\phi(x)$).
Distinguishing these last terms in $\phi(x)$ in Eq.~(\ref{secpezz}) from the 
others, one can write:  
\begin{equation} 
\label{f-f}
\left< \hat{f}_{i-1}(x)+\hat{f}_{i+1}(x)\right>= 
A(x_c)\phi(x)+B(x,x_c)\,, 
\end{equation}
where $A(x_c)$ and $B(x,x_c)$ are positive functions, and $B(x,x_c)=B(x_c,x_c)$
for $x>x_c$, because of Eq.~(\ref{refresh2}).

In principle, we should evaluate $\left<m_{i}(x)\right>$ applying the rule 
of conditional probability in analogy with Eq.(\ref{refresh2}), and then 
performing the average as in Eq.~(\ref{secpezz}). 
However, this would reduce Eq.(\ref{bala2}) to a trivial identity. 
For this reason, in order to evaluate $\left<m_{i}(x)\right>$, we adopt 
another strategy in the optics of the point (i).
We impose that $x_i$ is the minimal fitness in the whole system at a certain 
time-step $T\gg N$ from the actual beginning of the dynamics.
Consequently, we apply the RTS algorithm from the first time-step 
\cite{matteo} of the dynamics up to $T$ (in this case we have to consider 
all the system sites).
In the end, we consider the average $\left<...\right>$
over all the possible histories up to time $T$.
Since for $T\gg N$ the system must be in the stationary state \cite{rev}, 
then, after the average, this description must be consistent with the 
previous one where 
the system was considered directly in the stationary state.
In order to simplify the mathematical treatment, we use
the formula obtained for the generalized $\beta$-BS model \cite{IP-T}, 
and we take the limit 
$\beta \rightarrow \infty$ at the end of calculations:
\begin{equation} 
\langle m_{i}(x)\rangle=N\phi(x)\theta(x_c-x)\;, 
\label{m-aver}
\end{equation} 
where $x_c$ is defined by the relation 
$e^{-\beta x_c}=\int_0^1dx\, \phi(x) e^{-\beta x}$ with $\beta\gg 1$, which 
gives also a deeper meaning to $x_c$ \cite{IP-T}. 
Note that in this contest the presence of a finite threshold $x_c$ arises 
naturally, without external assumption. 
In order to obtain Eq.~(\ref{m-aver}), we used the central limit theorem 
for $N\gg 1$ \cite{IP-T} and we made the following approximation\cite{pinco}:
\[\left<\prod_{j=1}^N f_{j,T}(x_j)\right>\simeq \prod_{j=1}^N\left<f_{j,T}(x_j)
\right>= \prod_{j=1}^N\phi(x_j)\,.\]
Introducing Eqs.~(\ref{f-f}) and (\ref{m-aver}) in Eq.~(\ref{bala2}), 
we obtain:
\begin{equation} 
\label{phi}
\phi(x)=\frac{3-B(x,x_c)}{A(x_c)+N\theta(x_c-x)}\,. 
\end{equation}
Note that, because of the behavior of 
$B(x, x_c)$, $\phi(x)$ is constant for $x>x_c$, while for $x<x_c$ 
is of order $1/N$.
There is only one value of $x_c$ for which $\phi(x)$ is 
normalized (i.e.  $\int_0^1\phi(x)dx=1$). 
This value gives the value $x_c$ of the BS threshold.
 
In principle, one should consider the contribution to
Eq.~(\ref{secpezz}) coming from any
length $t$ in the avalanche tree (Fig.\ref{fig1}).
By stopping the sums at a value $t=n$, we obtain an $n$-order 
approximation.
We have evaluated the probability of the paths, through RTS,  
up to $n=7$ by using a computer program for numerical integration.
\begin{figure} 
\centerline{\psfig{file=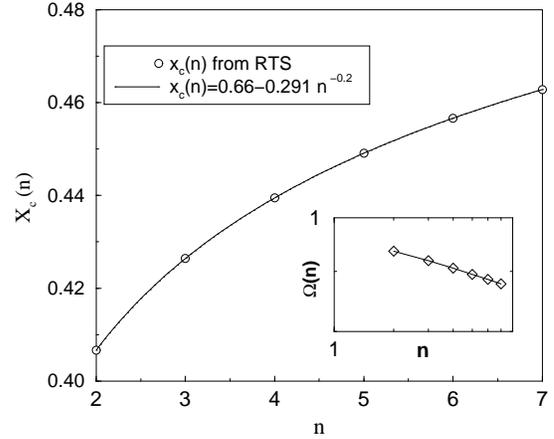,height=6cm,angle=0}} 
\caption{Empty points represent the values of $x_c$ 
obtained by the application of the RTS algorithm from $n=2$ to $n=7$.
The continuous line represents the fit curve $x_c(n)=0.66-ax^b$ with
$a=0.291\pm 0.003$ and $b=0.20\pm 0.03$.
The insert shows the behavior of $\Omega(n)$ up to $n=7$. Assuming 
$\Omega(n)\sim n^{-\tau+1}$
as a good approximation also for small $n$, one finds $\tau\simeq 1.05$.} 
\label{fig2} 
\end{figure} 
The results for $x_c$, for $n=2$ up to $n=7$, are reported in Fig.\ref{fig2}:
The best evaluation is $x_c(n=7)\simeq 0.465$, much better than the mean-field
result $x_c=1/3$, but still quite far from the above reported numerical value. 
However, we verified that this behavior is compatible with an asymptotic value
$x_c(n\rightarrow \infty)\simeq 0.66$. 
To this aim, we made a fit with the simplest possible 
function $x_c(n)=0.66-ax^b$ (Fig.\ref{fig2}) which
is well compatible with the given asymptotic value. 
The fit vaues are $a=0.291\pm 0.003$ and $b=0.20\pm 0.03$. 
This small value of $b$ 
is due to the fact that the avalanche size distribution $P(s)$ \cite{rev}
is characterized by a small exponent ($\tau\simeq 1.07$), 
henceforth all the sizes $s$ are important for statistics.
One can use $x_c(n\rightarrow \infty)$ to 
evaluate both the avalanche exponent $\tau$ and the average minimum 
distribution 
$m(x)\equiv \left<m_i(x)\right>$.
The exponent $\tau$ can be found studying, using 
$x_c(n\rightarrow\infty)=0.66$ 
in the Eqs.~(\ref{prob}) and (\ref{refresh2}), the behavior of
$\Omega(t)=\sum_{{\cal C}(t)} W({\cal C}(t))$ 
for $t$ ranging from $1$ to the maximal $n$
possible. Indeed  $\Omega(t)$ is proportional to
the probability that the avalanche lasts at least $t$ time-steps.
In the scaling regime, $\Omega(t)\sim t^{-\tau+1}$. 
Making this hypothesis, the result for $n=7$ is
$\tau(n=7)\simeq 1.05$ (see insert in Fig.\ref{fig2}), 
which is in agreement with the known numerical value.
Clearly, the hypothesis of scaling regime for $n=7$ is a strong hypothesis, 
but it is based on the fact that for Invasion Percolation, 
which has similar dynamical rules, it has been shown \cite{euro1} that 
the microscopical dynamical rules are already the scale invariant ones. 
Finally, we can obtain an approximation, of the distribution
of the minimal fitness $m(x)$ in the stationary state considering the 
Eqs.~(\ref{m-aver})
and (\ref{phi}), where the latter one is evaluated using RTS up to $n=7$, 
but using $x_c(n\rightarrow\infty)$ in the RTS calculations. 
\begin{figure}
\centerline{\psfig{file=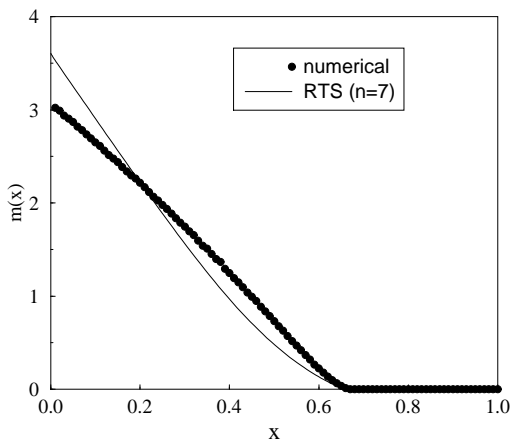,height=6cm,angle=-90}}
\caption{The continuous line gives the theoretical stationary distribution $m(x)$ of 
the minimal fitnesses where $n=7$ and assuming 
$x_c(n\rightarrow\infty)=0.66$.
The points represent the numerical behavior evaluated in extensive simulations.} 
\label{fig3} 
\end{figure} 
Imposing normalization, we obtain the function $m(x)$
reported in Fig.\ref{fig3}. In the same figure this result is compared with the 
known asymptotic numerical distribution of minimal fitnesses \cite{bs}.
The agreement is again quite good, considering the strong approximation
in truncating the sums of Eq.(\ref{secpezz}) at $n=7$.
Finally, it is worth to note that our method include also the mean-field results
for BS. We recall that \cite{derr} in the mean-field version of the model, when
a site $i$ is selected, the other two sites refreshed are not the fnn
sites,
but two randomly chosen system sites. This means that in Eq.(\ref{bala2})
one has to use $\left<f_{i-1}(x)\right>=\left<f_{i+1}(x)\right>=\phi(x)$.
Henceforth, the asymptotic histogram becomes:
\begin{equation}
\phi(x)=\frac{3}{2+N\theta(x_c-x)}\;,
\label{isto-mean}
\end{equation}
that is $\phi(x)=O(1/N)$ for $x<x_c$ and $\phi(x)=3/2$ for $x>x_c$.
Normalizing $\phi(x)$, one gets $x_c=1/3$, which is 
the above cited mean-field result.

In conclusion, this paper presents a kind of perturbative 
approach to the BS 
model, based on the probabilistic framework called Run Time Statistics. 
Through this approach, we compute
the self-organized threshold $x_c$, the avalanche exponent $\tau$, and
the stationary distribution of minimal fitnesses $m(x)$.
In principle one could also obtain in this way the exponent $\mu$ characterizing the 
sites covered by an avalanche.
However, since we can reach at the best the $7^th$ time step of growth, this results in a very poor statistics
for the number of points covered by the paths.
Authors acknowledge the support of the EU Network 
ERBFMRXCT980183.



\begin{thebibliography}{99}
\bibitem{pinco} Note that, in this case, the functions $f_{j,t}(x)$ should 
be obtained by applying the above mentioned strategy (i) following the 
dynamics from the real beginning and not directly in the stationary state.
\end{thebibliography}
\end{document}